\documentclass[aps,reprint,groupedaddress]{revtex4-1}

\usepackage{graphicx}

\begin{document}



\title{Epitaxy and Structural Properties of (V,Bi,Sb)$_2$Te$_3$ Layers Exhibiting the Quantum Anomalous Hall Effect} 



\author{M. Winnerlein}
\thanks{M. Winnerlein and S. Schreyeck contributed equally to this work.}
\author{S. Schreyeck}
\thanks{M. Winnerlein and S. Schreyeck contributed equally to this work.}
\author{S. Grauer}
\author{S. Rosenberger}
\author{K. M. Fijalkowski}
\author{C. Gould}
\author{K. Brunner}
\email{Karl.Brunner@physik.uni-wuerzburg.de}
\author{L. W. Molenkamp}


\affiliation{Physikalisches Institut, Experimentelle Physik III, Universit\"at W\"urzburg, Am Hubland, D-97074 W\"urzburg, Germany}

\date{\today}

\begin{abstract}
The influence of Sb content, substrate type and cap layers on the quantum anomalous Hall effect observed in V-doped (Bi,Sb)$_2$Te$_3$ magnetic topological insulators is investigated. Thin layers showing excellent quantization are reproducibly deposited by molecular beam epitaxy at growth conditions effecting a compromise between controlled layer properties and high crystalline quality. The Sb content can be reliably determined from the in-plane lattice constant measured by X-ray diffraction, even in thin layers. This is the main layer parameter to be optimized in order to approach charge neutrality. Within a narrow range at about 80\% Sb content, the Hall resistivity shows a maximum of about 10 k$\Omega$ at 4 K and quantizes at mK temperatures. Under these conditions, thin layers grown on Si(111) or InP(111) and with or without a Te cap exhibit quantization. The quantization persists independently of the interfaces between cap, layer and substrate, the limited crystalline quality, and the degradation of the layer proving the robustness of the quantum anomalous Hall effect. 

\end{abstract}

\pacs{}

\maketitle 


A quantum anomalous Hall effect (QAHE), in which a Hall plateau with a resistance of $h/e^2$ can be observed even in the absence of a magnetic field, was predicted to occur in ferromagnetically doped topological insulators (TIs) \cite{Onoda2003,Liu2008,Yu2010,Nomura2011}. The effect was first observed in 2013 in a 5 nm Cr$_{0.15}$(Bi$_{0.1}$Sb$_{0.9}$)$_{1.85}$Te$_3$ layer without a cap layer, and grown on a SrTiO$_3$ substrate \cite{Chang2013}. The QAHE has since been reproduced in such tetradymite-type layers with different layer thicknesses, layer compositions, magnetic dopants (Cr or V), with/without a cap layer and on various substrates \cite{Chang2014,Checkelsky2014,Kou2014,Bestwick2015,Chang2015,Grauer2015}. Perfect quantization was observed in Cr-doped (Bi,Sb)$_2$Te$_3$ layers capped by an AlO$_x$ layer on GaAs substrates, and V-doped (Bi,Sb)$_2$Te$_3$ layers capped by Te on SrTiO$_3$ substrates and Si(111)  substrates \cite{Chang2014,Bestwick2015,Chang2015,Grauer2015}. Many authors point out that the layer properties have to be optimized to observe the QAHE, but details of the epitaxial growth conditions, the structural layer properties and the applied characterization techniques are sparse. 

An apparent requirement for observing the QAHE is the optimization of Sb content in order to approach charge neutrality by compensation of n- and p-type intrinsic point defects. Further, an applied gate bias is needed to tune the Fermi level into the bandgap of topological surface states (TSSs) caused by magnetization \cite{Chang2013}. The layer thickness, ranging from about 4 to 10 nm in the literature, likely affects the hybridization of the TSSs and possibly also the mechanisms leading to the QAHE. Theoretical predictions  and experiments both reveal a degradation of the surface states due to oxidation in Bi$_2$Te$_3$ layers and shifts of the Fermi level in epitaxial tetradymite layers due to adsorbates \cite{Wang2012,Kong2011,Park2015,Maass2014}. The influence of the interfaces between the layer and the substrate, as well as between the layer and either an AlO$_x$ or Te cap layer applied for surface protection on magnetotransport studies revealing the QAHE is still not clear. A high structural quality of the layers has been claimed to be a prerequisite of QAHE, while epitaxial tetradymite layers in general suffer from rotational twinning, translational domain boundaries, screw dislocations and quintuple layer (QL) surface steps \cite{Tarakina2012,Tarakina2014}. The coincidence of the QAHE and contributions of superparamagnetic phases observed in magnetotransport in the nominally ferromagnetic phase of such layers as well as a multi-domain network during the magnetization reversal process suggest that the layers are inhomogeneous in magnetic as well as structural properties \cite{Grauer2015,Kou2015}.

In this letter, we study the impact of growth conditions during molecular beam epitaxy (MBE) and the resulting structural properties of V$_{0.1}$(Bi$_{1-x}$Sb$_x$)$_{1.9}$Te$_3$ layers on magnetotransport and the observation of the QAHE. The layers are of a constant thickness of about 9 nm, deposited on Si(111) and InP(111) substrates, with and without a Te cap layer and of varied Sb content $x$. For growth conditions similar to those reported in literature and an Sb content of about $x = 0.8$, Hall resistivities close to or at the quantized value $h/e^2$ are observed in magnetotransport at mK temperatures \cite{Chang2013,Chang2014,Bestwick2015,Chang2015}. The results reveal no significant influence of the limited structural layer quality, the substrate or the cap layer. The Sb content required to obtain charge neutrality of the layer is the most important parameter for the observation of the QAHE. Even in thin layers, the Sb content can be reliably determined from the lateral lattice constant $a$, which is rather insensitive to layer thickness, substrate temperature and V content. MBE growth parameters such as substrate and V cell temperature do affect the layer growth rate and the Sb content. Consequently, the Sb content has to be analyzed and optimized for each set of MBE growth parameters.

MBE growth conditions and their impact on layer properties will be described in detail in the next section. The morphology, thickness and crystallinity of layers were characterized by atomic force microscopy (AFM) in ambient conditions, X-ray reflectivity (XRR) measurements and X-ray diffraction (XRD) $\theta-2\theta$ and azimuthal $\phi$ scans. Hall bars with top gates were produced by optical lithography, as described elsewhere \cite{Grauer2015}. Magnetotransport studies were performed either in $^4$He ($T=4$ K) or a $^3$He-$^4$He dilution refrigerator cryostat ($T\leq 50$ mK) by low-frequency lock-in techniques.

\begin{figure}
\includegraphics[width=0.9\linewidth]{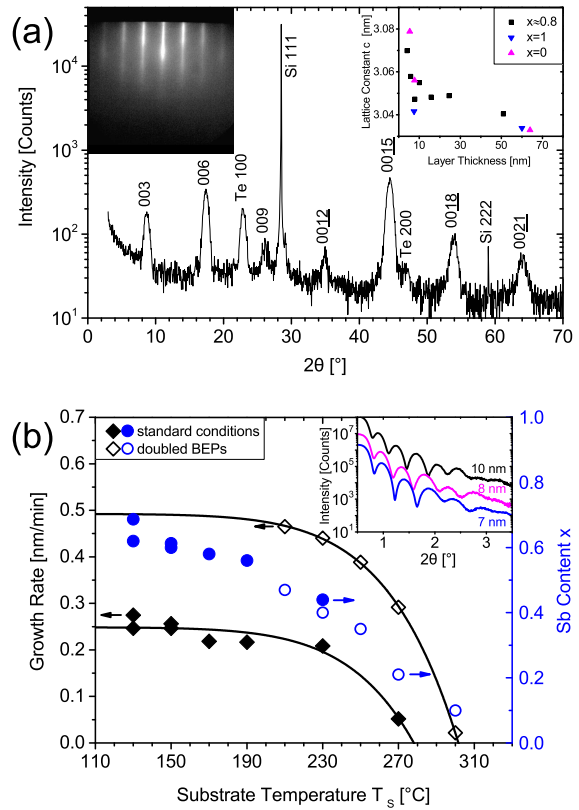}
\caption{\label{fig_1} (a) $\theta$-2$\theta$ XRD scans of a 9 nm thick V$_{0.1}$(Bi$_{0.21}$Sb$_{0.79}$)$_{1.9}$Te$_3$ layer with 10 nm Te cap on Si(111) (sample 2). The left hand inset shows a RHEED pattern from the surface of a representative layer. In the right hand inset the lattice constant $c$ is shown as a function of layer thickness for $x = 0$ (magenta), $x = 1$ (blue) and $x \approx 0.8$ (black).\\
(b) Dependence of the layer growth rate (black) and the Sb content $x$ (blue) on the substrate temperature $T_s$ for standard growth conditions (full symbols) and doubled BEPs of Te, Bi, and Sb  (empty symbols). The inset shows XRR scans of a 7 nm (blue), 8 nm (magenta) and 10 nm (black) thick layer with 10 nm Te cap (offset by a factor of 10 each for clarity).}
\end{figure}

Prior to MBE growth, the substrates are dipped in 50\% HF solution and immediately loaded into the MBE system. The MBE is equipped with thermal effusion cells and the substrate is heated by thermal radiation. The substrate temperature $T_s$ is monitored by a thermocouple on the backside of a PBN diffusor plate on which the 2-inch substrates are clamped. The H-passivated Si(111) substrates are ramped directly to a specified $T_s$, while semi-insulating Fe:InP(111)B substrates are heated to about $640^{\circ}$C in Te atmosphere in order to clean and smoothen the surface and subsequently cooled to a given $T_s$ \cite{Tarakina2014}. The beam equivalent pressure (BEP) of molecular beams is measured by a Bayard Alpert ion gauge at the substrate position. Under our standard conditions, the BEP of Te is $2.7\times10^{-7}$ mbar and the total BEP of group-V materials is about $7\times10^{-8}$ mbar. These Te-rich conditions result in a layer growth rate of $0.2 - 0.3$ nm/min at T$_s=190^{\circ}$. The V cell is kept at $1400^{\circ}$C providing a constant flux, which was regularly checked by measuring the thickness of pure V reference layers by XRR. The V content $z$ was calibrated by energy dispersive X-ray spectroscopy (EDX) of a bulk V$_{0.1}$(Bi$_{0.24}$Sb$_{0.76}$)$_{1.9}$Te$_3$ reference layer. This allows us to deduce the V content $z$ of the V$_z$(Bi$_{1-x}$Sb$_x$)$_{2-z}$Te$_3$  layers from their overall growth rate determined from the growth duration and layer thickness. Reflection high-energy electron diffraction (RHEED) shows streaks during growth as shown in the left hand inset in Fig. \ref{fig_1} (a), suggesting a smooth (V,Bi,Sb)$_2$Te$_3$ layer. After the growth of the layer, an optional 10 nm Te cap layer is deposited \textit{in-situ} at about $T_s=20^{\circ}$C. RHEED on this cap layer shows growth of amorphous Te for the first 3 nm followed by 7 nm crystalline Te.

Fig. \ref{fig_1} (a) shows a $\theta-2\theta$ XRD scan of a representative 9 nm thick V$_{0.1}$(Bi,Sb)$_{1.9}$Te$_3$ layer grown at T$_s$=190$^{\circ}$C on Si(111) with a 10 nm Te cap. The peaks are identified as the (00$\underline{3n}$) reflections of the layer as well as peaks from the Si substrate and the Te cap layer. The parallelism of (001) layer planes to (111) planes of the substrate is confirmed and no phases other than (V,Bi,Sb)$_2$Te$_3$ are detected. The widths of the layer peaks correspond to the layer thickness of about 9 nm. The position of the (006) peak is used to determine the out-of-plane lattice constant $c$. In the right hand inset of Fig. \ref{fig_1} (a), the lattice constant $c$ is shown as a function of layer thickness for Sb contents $x = 0$, $x \approx 0.8$ and $x = 1$. For layers of about 60 nm thickness, nearly no difference between the lattice constants $c$  for $x = 0$ and $x = 1$ is observed, which is consistent with the literature \cite{Smith1962,Wyckoff1964,Anderson1974,NAKAJIMA1963}. With decreasing layer thickness, the layers show a trend towards higher values and considerable variations of $c$. We also observed a trend that $c$ increases by up to $0.3$ \AA\ with decreasing substrate temperature in the range $T_s=130 - 270^{\circ}$C (not shown here). The variations of $c$, both with layer thickness and $T_s$, may be assigned to changes in length of the weak van-der-Waals bonds between neighboring QLs and at the interface to the substrate due to intercalates caused by non-ideal growth conditions \cite{Tarakina2014}. This leads to the conclusion that the lattice constant $c$ is not suitable to determine the Sb content.

The in-plane lattice constant $a$ is determined from the symmetric (006) and asymmetric $\{$015$\}$ reflections. It changes significantly with Sb content from $a = 4.383$ \AA\ in V$_{0.1}$Bi$_{1.9}$Te$_3$ to $a = 4.258$ \AA\ in V$_{0.1}$Sb$_{1.9}$Te$_3$ layers. Both values are about $0.01$ \AA\ below those of epitaxial layers grown without V, which is close to our accuracy limit of about $\pm 0.005$ \AA\ in determining $a$. The lattice constants at such a low V doping level are in the range of literature values for pure bulk Bi$_2$Te$_3$ and Sb$_2$Te$_3$ \cite{Smith1962,Wyckoff1964,Anderson1974,NAKAJIMA1963}. We further observed that the lattice constant $a$, in contrast to $c$, is not significantly affected by layer thickness or substrate temperature. This insensitivity on growth conditions, as well as the shifts of $a$ with layer composition can be explained by considering the strong chemical bonds within the QLs, which rigorously define the lattice constant $a$. Thus, we assume the validity of Vegard's law and determine the Sb content $x$ of V$_{z}$(Bi$_{1-x}$Sb$_x$)$_{1-z}$Te$_3$ layers by interpolation between the measured lattice constants $a$ of reference samples with $x=0$ and $x=1$ grown at preferably identical conditions ($z = 0.1$). The Sb content determined by XRD via $a$ has been compared to an elemental analysis of layer composition by EDX and secondary ion mass spectrometry (SIMS). We have used a V$_{0.1}$(Bi$_{1-x}$Sb$_x$)$_{1.9}$Te$_3$ layer with a larger thickness of 51 nm to increase the accuracy of the applied techniques. The XRD method yields $x = 0.76$, while EDX results in $x = 0.79$ and SIMS in $x = 0.71$. The values agree within an accuracy of $\pm 0.05$ for the Sb content $x$ determined by XRD, which is a method applicable even for the thin layers required for the QAHE.

In Fig. \ref{fig_1} (b) the growth rate and Sb content of V$_{0.1}$(Bi$_{1-x}$Sb$_x$)$_{1.9}$Te$_3$ layers as a function of $T_s$ is shown for standard MBE conditions (full symbols) and for doubled fluxes of Te, Bi, and Sb (open symbols). The layer thickness and thus the growth rate are determined from the fringe positions in XRR, which are sensitive even for small differences below 1 QL in layer thickness (inset of Fig. \ref{fig_1} (b)). At our standard conditions and low $T_s$ ($\leq 190^{\circ}$C) the growth rate as well as the Sb content are nearly independent from $T_s$. The Sb/Bi content ratio agrees to within $\pm 15\%$ with the calculated Sb/Bi atomic flux ratio. Here we assume a flux of Bi$_2$ molecules each offering two Bi atoms and of Sb$_4$ molecules, which dissociate and offer also 2 atoms in a surface reaction process as described for As$_4$ \cite{Wood1982,Joyce1993,Joyce1975}. The growth rate, Sb and V content of layers are thus well controlled. 

At higher $T_s$ ($> 190^{\circ}$C), the observed reduction in growth rate and Sb content suggests an increasing thermal desorption of atoms, majoritarily of Sb. In this regime, the layer properties depend strongly on $T_s$ and the V cell temperature. It is worth noting here, that heat radiation emitted from the hot V cell appears to lead to an actual substrate temperature considerably above the nominal value $T_s$ given by the thermocouple. We estimate the difference to be about $30^{\circ}$C. For  $T_s \leq 230^{\circ}$C, a doubling of the BEPs of Te, Bi and Sb results in a doubling of the growth rate, as expected. At higher $T_s > 230^{\circ}$C, the growth rate is more than doubled when doubling the BEPs. The growth rate and the Sb content in this regime depend strongly on growth parameters and are hard to control due to Sb desorption.

We chose $T_s = 190^{\circ}$C and standard BEPs as a compromise for realizing layers with reasonably good structural quality as well as reproducibility in thickness and composition. This stable growth window allows us to systematically vary the layer properties in order to study the requirements for realizing the QAHE.

\begin{figure}
\includegraphics[width=0.9\linewidth]{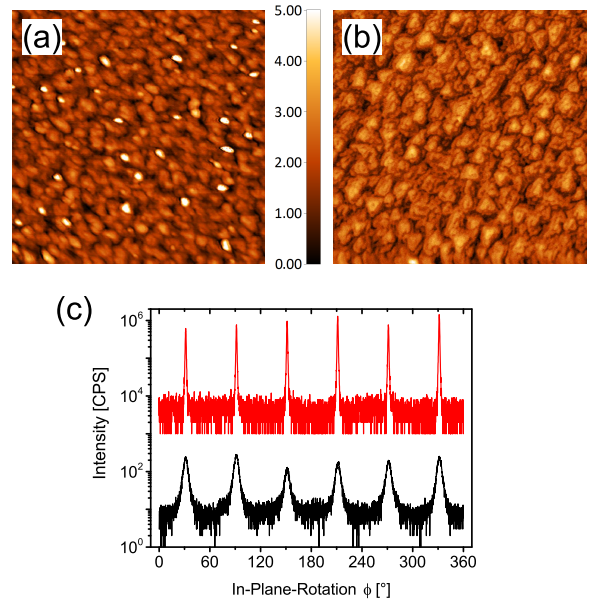}
\caption{\label{fig_2} (a) and (b) AFM images ($2 \times 2 \, \mu m ^2$, 5 nm z scale) showing the surface of 10 nm thick V$_{0.1}$(Bi$_{0.21}$Sb$_{0.79}$)$_{1.9}$Te$_3$ layers without Te cap on Si(111) (sample 1) and on InP(111) respectively. (c) In-plane rotation $\phi$ scans of $\{$015$\}$ reflections from 9 nm thick layers with Te cap on Si (black, sample 2) and on InP (red, sample 3) (vertically offset for clarity).}
\end{figure}

Next we discuss the morphology and crystalline quality of V$_{0.1}$(Bi$_{1-x}$Sb$_{x}$)$_{1.9}$Te$_3$ layers with a thickness of about 9 nm and $x \approx 0.8$, which show the QAHE (samples 1, 2 and 3).  Fig. \ref{fig_2}(a) and (b) present AFM images of layers without Te cap grown under the same growth conditions on Si(111) (sample 1) and on InP(111). The obtained RMS roughness of each layer is 0.7 nm. The layers consist of islands with an average diameter of about 100 nm on Si and 140 nm on InP. The islands on InP tend to be larger, more interconnected, and with wider QL terraces at the surface. The layer on Si(111) reveals a few islands up to 12 nm height and deep trenches between islands. All studied layers adequate for observing the QAHE are quite rough compared, for example, to layers grown at higher $T_s = 250^{\circ}$C with larger islands merged on a $\mu$m scale (not shown) but with poorly controlled composition.

The crystalline quality is studied by XRD $\phi$ scans of the asymmetric $\{$015$\}$ reflections from Te capped 9 nm thick V$_{0.1}$(Bi$_{0.21}$Sb$_{0.79}$)$_{1.9}$Te$_3$ layers on Si (sample 2) and InP (sample 3), see Fig. \ref{fig_2}(c). On both types of substrate six peaks are observed, which are a clear signature of rotational twinning in the layers \cite{Schreyeck2013}. The FWHM is as large as 5.4$^{\circ}$ on Si and only 1.0$^{\circ}$ on InP.  The in-plane orientation of the islands is significantly better on InP than on Si, which is attributed to the smaller lattice mismatch of InP \cite{Tarakina2012}.

\begin{figure}
\includegraphics[width=0.8\linewidth]{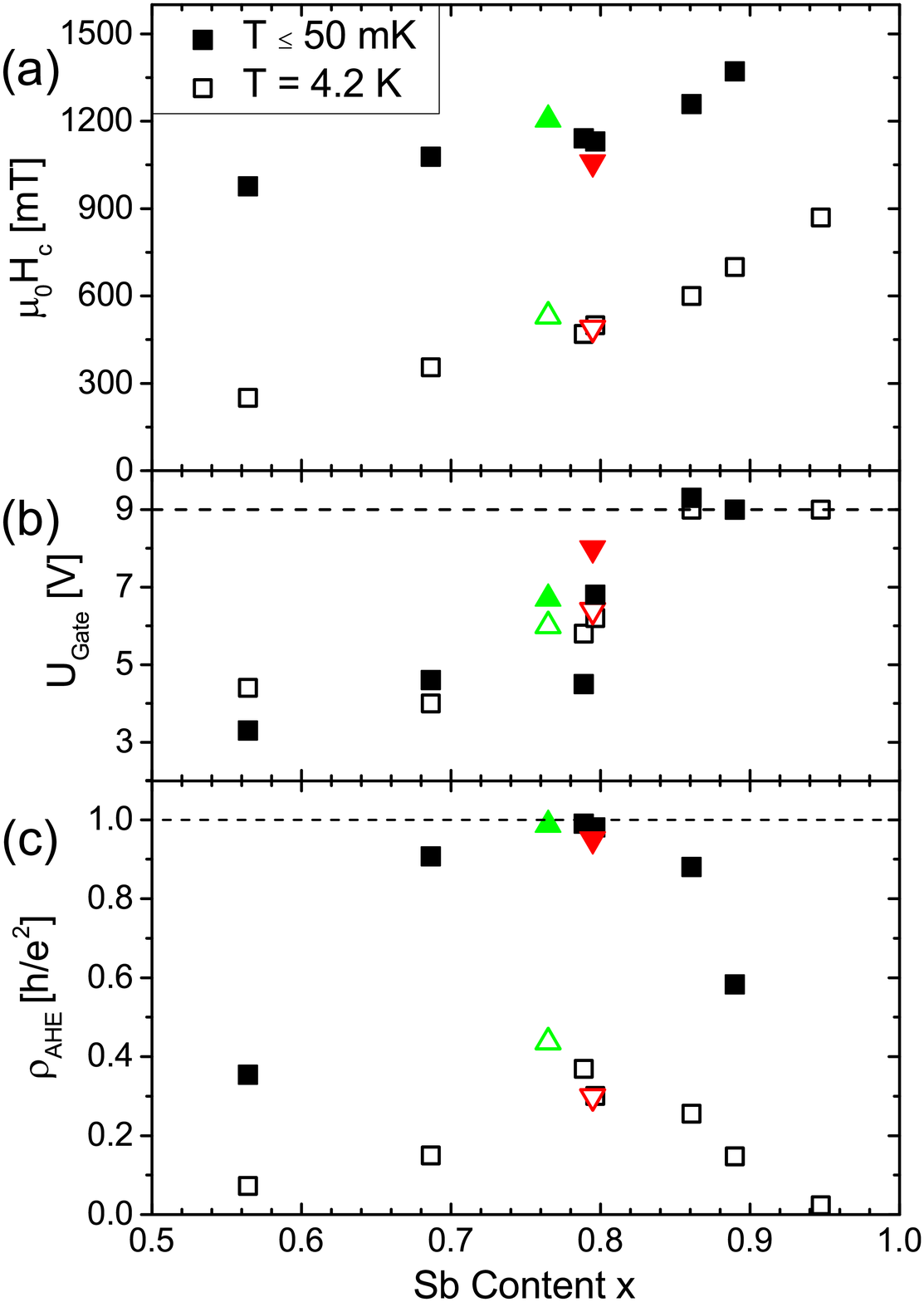}
\caption{\label{fig_3} Magnetotransport properties as a function of Sb content $x$ in different layers measured at $T = 4$ K (empty symbols) and $T \leq 50$ mK (full symbols). Te capped layers on Si (including sample 2) are shown in black, sample 1 (no cap, on Si) in green and sample 3 (with Te cap, on InP) in red:\\
(a) Coercive field $H_C$, (b) Gate voltage $U_{Gate}$ applied for observing maximal Hall resistivity $\rho_{AHE}$ up to the limit of 9 V due to gate breakdown, and (c) Anomalous Hall resistivity $\rho_{AHE}$ measured at applied gate voltage $U_{Gate}$.}
\end{figure} 

\begin{figure}
\includegraphics[width=0.8\linewidth]{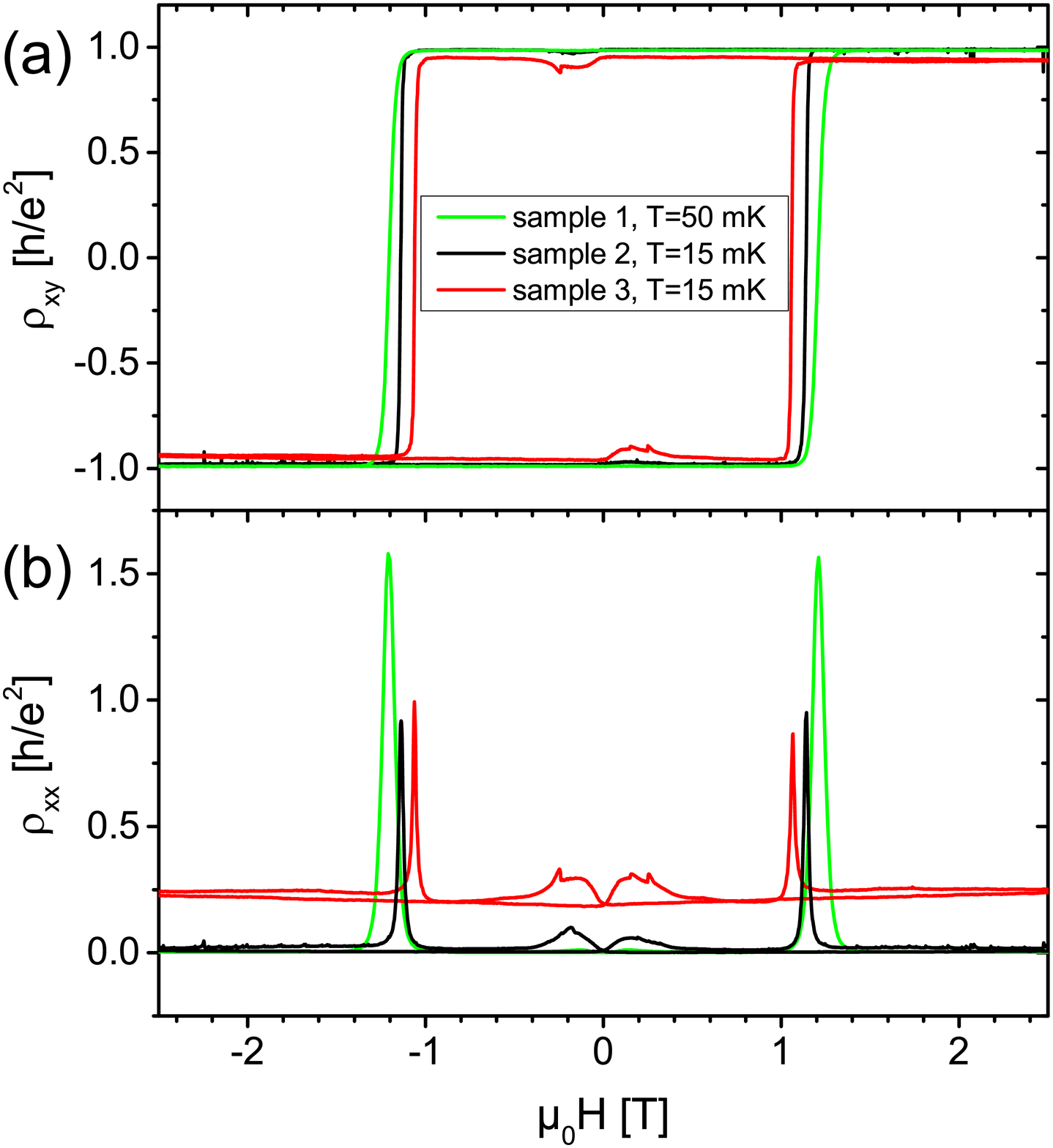}
\caption{\label{fig_4} (a) The Hall resistivity $\rho_{xy}$ and (b) longitudinal resistivity $\rho_{xx}$ of sample 1 (green), sample 2 (black) and sample 3 (red). All samples show the QAHE and a vanishing $\rho_{xx}$, except for sample 3, the V$_{0.1}$(Bi$_{0.21}$Sb$_{0.79}$)$_{1.9}$Te$_3$ layer with Te cap on InP.}
\end{figure}

In the following, the influence of layer properties on magnetotransport and on the observation of the QAHE is studied.  Fig. \ref{fig_3} (a) displays the coercive field depending on the Sb content $x$. The coercive field increases monotonically with increasing $x$ both at $T = 4$ K (empty symbols) and $T < 50$ mK (full symbols). Fig. \ref{fig_3} (b) shows the gate voltage that maximizes the Hall resistivity, which itself is plotted in Fig. \ref{fig_3} (c). The gate voltage for maximal Hall resistivity tends to higher voltages with increasing $x$ until $x>0.86$, where $U_{Gate} = 9$ V is limited by the breakdown voltage of the gate. The anomalous Hall resistivity reaches a maximum at Sb contents of about $x = 0.8$ for both temperatures. This suggests that the measurements necessary to optimize the Sb content can be conducted at $T = 4$ K. At $T < 50$ mK the QAHE is observed in layers with an Sb content in a narrow range of about $x = 0.76 - 0.79$. This is irrespective of whether the layer is without Te cap on Si (sample 1), with Te cap on Si (sample 2) or with Te cap on InP (sample 3).
Note that at $x = 0.86$ and $T < 50$ mK the Hall resistivity achieves a maximum within our gate voltage range, but does not reach the quantized value. This shows that the decline of the Hall resistivity at such a high Sb content is not due to the limitation in gate voltage. The high sensitivity of the AH resistivity on Sb content can be tentatively assigned to p-type intrinsic defects such as Sb vacancies and Sb$_{Te}$ antisites in Sb$_2$Te$_3$ which compensate and eventually outweigh the n-type intrinsic defects like Te$_{Bi}$ antisites in Bi$_2$Te$_3$ with increasing $x$ \cite{Jiang2012,Chen2009}. Consistently, a change in sign of the slope of the Hall resistivity at high magnetic fields is observed between $x = 0.80$ to $x = 0.89$ at 4 K as shown in the supplement \cite{supplement}. The QAHE is observed close to this charge neutrality point \cite{Yu2010}.

The measured longitudinal and Hall resistivity data of sample 1, 2 and 3 are compared in Fig. \ref{fig_4}. Both samples on Si show perfect quantization in the Hall resistivity as well as zero longitudinal resistivity within measurement accuracy. Note that the uncapped sample was under ambient conditions for several months, i.e. sufficient time for a possible layer degradation, before the processing and measurements were performed. This proves that a cap layer is not necessary for the observation of the QAHE. 
Sample 3 on InP substrate shows a reduced Hall resistivity of $0.95~h/e^2$ and a finite longitudinal resistivity of about 5.2 k$\Omega$. Remarkably, the layer with the higher crystalline quality compared to samples 1 and 2 deviates from quantization. A reason for this might be parasitic conduction by bulk-like carriers which remain delocalized in the more homogenous layer at mK temperatures. Given our limited accuracy ($\pm 5\%$) in determining the Sb content $x$ and the obviously narrow range required for QAHE, however, we can not exclude a slight deviation in the actual $x$ of sample 3 as an alternative explanation. The observed features at low fields $< 0.6$ T were already reported and discussed elsewhere \cite{Grauer2015}. The smaller feature amplitude for sample 1 is likely not due to sample specifics but rather to the use of a different measurement setup with less RF filtering and increased effective sample temperature.

In summary, V$_{0.1}$(Bi,Sb)$_{1.9}$Te$_3$ layers are reproducibly grown by MBE on different substrates. Layers grown under the same growth conditions on Si(111) and InP(111) substrates with/without a cap layer exhibit the QAHE. The main requirement for quantization is an Sb content close to about $x = 0.8$ in order to approach charge neutrality of the layer, while its interfaces and rather poor crystalline quality are of minor importance. The Sb content even in thin films, is quite accurately determined from the in-plane lattice constant $a$ measured by XRD.

We gratefully acknowledge the financial support of the EU ERC-AG Program (project 3-TOP), the DFG through SFB 1170 "ToCoTronics" and the Leibniz Program.




%

\end{document}